\documentclass[fleqn,10pt]{wlscirep}%
\usepackage{amsfonts}
\usepackage{amsmath}
\usepackage{graphicx,color,psfrag,amssymb,amsmath,url}
\usepackage{float}
\usepackage{setspace}
\usepackage{url}
\usepackage{epstopdf}
\usepackage{amssymb}
\usepackage{graphicx}%
\providecommand{\U}[1]{\protect\rule{.1in}{.1in}}
\newtheorem{theorem}{Theorem}
\newtheorem{proposition}[theorem]{Proposition}

\begin{abstract}
When applied to matter and antimatter states, the Pauli master equation
(PME) may have two forms: time-symmetric, which is conventional, and
time-antisymmetric, which is suggested in the present work. The symmetric
and antisymmetric forms correspond to symmetric and antisymmetric extensions
of thermodynamics from matter to antimatter --- this is demonstrated by
proving the corresponding H-theorem. The two forms are based on the
thermodynamic similarity of matter and antimatter and differ only in the
directions of thermodynamic time for matter and antimatter (the same in the
time-symmetric case and the opposite in the time-antisymmetric case). We
demonstrate that, while the symmetric form of PME predicts an equi-balance
between matter and antimatter, the antisymmetric form of PME favours full
conversion of antimatter into matter. At this stage, it is impossible to
make an experimentally justified choice in favour of the symmetric or
antisymmetric versions of thermodynamics since we have no experience of
thermodynamic properties of macroscopic objects made of antimatter, but
experiments of this kind may become possible in the future. \
\end{abstract}
\begin{document}

\title{Symmetric and antisymmetric forms of the Pauli master equation \\ \Large{for interaction of matter and antimatter quantum states} \\ \large{\it Published: Nature.com - Scientific Reports 6, 29942 (2016)}}
\author{Klimenko A. Y.}

\affil[1]{The University of Queensland, SoMME, QLD 4072, Australia}

\flushbottom

\maketitle

\thispagestyle{empty}

\section*{Introduction}

Microscopic objects are governed by the equations of quantum mechanics and
involve both particles and antiparticles. These equations are time-reversible
and do not discriminate between the past and the future
\cite{Hawking-Penrose,Zeh2007}. Nonetheless, the macroscopic objects, which
are common in our day-to-day lives, are subject to the laws of thermodynamics
that are irreversible in time.\ The second law of thermodynamics predicts
irreversible increase in entropy and, thus, strongly and unambiguously
discriminates the directions of time \cite{PrigThermo,PriceBook,Zeh2007}: the
direction of the thermodynamic time points to the direction of entropy
increase. There is another apparent asymmetry in the Universe: macroscopic
objects are exclusively made of matter (particles). Antimatter, which is
formed by antiparticles in the same way as matter is formed by particles, is
theoretically possible but seems not to be present anywhere in the known
Universe\cite{PenroseBook}. While the microscopic properties of antiparticles
are generally well known, the fundamental conceptual problem that we are
facing is the extension of macroscopic properties from matter to antimatter.
We expect that properties of matter and antimatter are in some way similar,
but it appears that extension of the second law of thermodynamics from matter
to antimatter is not unique, allowing for two possible alternatives:
time-symmetric and time-antisymmetric \cite{KM2014}. The thermodynamic times
of matter and antimatter run in the same direction according to the former,
and in the opposite directions according to the latter. These two
possibilities are referred to in the paper as the symmetric and antisymmetric
versions of thermodynamics. These versions are mutually incompatible and,
since we do not have any experience with macroscopic antimatter, it is not
known which one of these versions is real.

One can expect that a quantum (or classical) system of a sufficiently large
dimension and complexity should display thermodynamic properties. The apparent
discrepancy between the irreversible equations of thermodynamics and
reversible equations of classical and quantum mechanics are mitigated by
so-called kinetic (or master) equations \cite{Zeh2007}. Equations of this type
are derived from the time-reversible equations of classical or quantum
mechanics, but necessarily involve assumptions discriminating the direction of
time. If entropy evolved by master/kinetic equations increases forward in
time, then these equations are consistent with thermodynamics. The important
statements demonstrating monotonic increase of entropy are traditionally
called H-theorems after the famous theorem by Ludwig Boltzmann
\cite{Boltzmann-book}, who demonstrated increase of entropy in gases under the
conditions of validity of the hypothesis of molecular chaos (the
\textit{Stosszahlansatz}). This hypothesis discriminates the direction of time
by assuming that the parameters of molecules are uncorrelated before (but not
after) their collisions. The gap between unitary (time-reversible) quantum
mechanics and thermodynamics was bridged by Wolfgang Pauli \cite{Pauli1928},
who derived a master equation by assuming decoherence of the states of a large
quantum system before (but not after) unitary interactions of the states takes
place. This class of equations, which is referred to as the Pauli master
equations (PME), is consistent with thermodynamics: it tends to increase
entropy and leads to microcanonical distributions. Demonstration of validity
of the corresponding H-theorem was one of the main goals of Pauli's
article\cite{Pauli1928}.

While there have been many attempts to improve or generalise PME
\cite{Zeh2007}, it seems that some of the more recent changes brought into our
understanding of PME have had more of a philosophical or methodological
character. While in the early days of quantum mechanics randomisation of
quantum phases was viewed as an additional assumption contaminating the
scientific rigor of quantum equations \cite{vanHove1954}, a more modern
treatment of this problem \cite{Joos1984,Zeh2007} is that decoherence is a
real physical process, whose exact mechanisms are not fully known at this
stage. This process primes the direction of thermodynamic time and causes
entropy increase. Joos \cite{Joos1984} noted that one of Pauli's
\cite{Pauli1928} remarks can be interpreted as pointing to environmental
interference. Smaller quantum systems are subject to decoherence due to
interference of the environment, while very rapid decoherence of larger
systems (i.e. macroscopic objects) is often seen as being indicative of the
presence of intrinsic mechanisms of decoherence \cite{Stamp2012}. While it
seems that both mechanisms of decoherence (i.e. environmental and intrinsic)
are possible and both are discussed in the literature \cite{Stamp2012}, we are
interested in the consequences of decoherence and do not dwell on its physical causes.

While the symmetric version of thermodynamics is conventionally implied in
publications, the possibility of the antisymmetric version raises a number of
questions. First, there must be a corresponding antisymmetric version of the
PME, which is to be derived from the conventional unitary equations of quantum
mechanics. This derivation endeavours to examine the link between macroscopic
and microscopic symmetries. Consistency of different versions of PME with the
corresponding versions of thermodynamics demands validity of the relevant
H-theorems. The question of properties of the derived equations, especially
whether PME statistically favours conversion of matter into antimatter or
antimatter into matter or is neutral with respect to this conversion, is of
particular interest.

\section*{A generic quantum system and perturbation analysis of its unitary
evolution}

\subsection*{System Hamiltonian}

While considering evolution of a generic quantum system, it is common to
distinguish two components in its Hamiltonian:
\begin{equation}
\mathbb{H}=\mathbb{H}_{0}+\mathbb{H}_{1}=\mathbb{H}_{0}+\lambda\mathbb{V}
\label{1Ham}%
\end{equation}
the leading time-independent component $\mathbb{H}_{0}$ determining the energy
eigenstates
\begin{equation}
\mathbb{H}_{0}\left\vert j\right\rangle =\varepsilon_{j}\left\vert
j\right\rangle \label{1ES}%
\end{equation}
and the disturbance $\mathbb{H}_{1}=\mathbb{H}_{1}(t)$ that characterises
interactions of these eigenstates $\left\vert j\right\rangle $ and may be
time-dependent \cite{Pauli1928,vanHove1954,Vliet1978,Zeh2007}. Different
states (i.e. eigenstates) may belong to the same energy levels $\varepsilon
_{j_{1}}=\varepsilon_{j_{2}},$ but these states are still marked by different
indices $j_{1}\neq j_{2}$. The characteristic time associated with the leading
Hamiltonian is much smaller than that of the interactions: $\tau_{0}%
\sim1/\left\vert \mathbb{H}_{0}\right\vert \ll\tau_{1}\sim1/\left\vert
\mathbb{H}_{1}\right\vert $ and a small parameter, $\lambda\equiv\tau_{0}%
/\tau_{1}\ll1,$ is used in (\ref{1Ham}) to explicitly reflect that
$\mathbb{H}_{0}\gg\mathbb{H}_{1}=\lambda\mathbb{V}.$ Here, $\left\vert
\mathbb{H}\right\vert $ denotes a norm estimate for the Hamiltonian
$\mathbb{H}$. The energy eigenstates form a complete orthogonal basis
$\left\langle j|k\right\rangle =\delta_{kj}$ where $\varepsilon_{j}$
represents energy of the $j^{\text{th}}$\ eigenstate. The number of
eigenstates, $2n$, is presumed large (eigenstates can be continuous). The
Hamiltonian disturbance $\mathbb{H}_{1}$ is responsible for interaction of the
eigenstates due to non-diagonal elements. If $\mathbb{H}_{1}$ depends on time,
the characteristic time of this dependence is assumed to be of order of
$\tau_{1}$. The diagonal elements of $\mathbb{H}_{1}$ are not of interest and
we put%
\begin{equation}
\left\langle j\middle|\mathbb{H}_{1}\middle|j\right\rangle =0
\end{equation}
Mathematically, this assumption simplifies the analysis but does not restrict
generality since the diagonal elements of $\mathbb{H}_{1}$ can always be
merged with the diagonal elements of $\mathbb{H}_{0}$.

\subsection*{Invariant properties of the system}

Two types of states are distinguished: the matter states and the antimatter
states. The matter states are indexed by $\tilde{j}$,$\tilde{k}=1,...,n$,
while the antimatter states are indexed by $\bar{j},\bar{k}=-1,...,-n$. The
indices $j$ and $k$ run over all $2n$ states. We also use the indicator
function $C_{j}=\pm1$ defined by
\begin{equation}
C_{\tilde{j}}=+1,\ \ \ \ \ \ \ \ \ C_{\bar{j}}=-1
\end{equation}
If $\left\langle +j\middle|\mathbb{H}_{1}\middle|-j\right\rangle \neq0,$ the
Hamiltonian (\ref{1Ham}) allows for conversion between the corresponding
matter states and antimatter states. The states $\left\vert +j\right\rangle $
and $\left\vert -j\right\rangle $ are presumed to represent the CP
transformations of each other
\begin{equation}
\widehat{\text{CP}}\left\vert j\right\rangle =\left\vert \widehat{\text{CP}%
}(j)\right\rangle =\alpha_{j}\left\vert -j\right\rangle
\end{equation}
although the physical coordinates and their parity (P) transformation, which
reverses the directions of these coordinates, are not explicitly considered
here. The CP transformation also involves the charge conjugation (C), which
changes particles into antiparticles and wise versa. The time reversal
operator (T) reverses the direction of time. The complex phase $\left\vert
\alpha_{j}\right\vert =1$ is subject to a number of physical constraints and,
in most cases, can be eliminated by incorporating the phase angle into the
states. Since our consideration is generic, we keep the phases $\alpha_{j},$
which, however, can be omitted as they do not affect our consideration and results.

Under conditions considered here, the CP- and CPT-invariant Hamiltonians
satisfy:
\begin{align}
\text{CP}  &  \text{:\ \ \ \ }\left\langle j\middle|\mathbb{H}%
\middle|k\right\rangle =\left\langle \widehat{\text{CP}}(j)\middle|\mathbb{H}%
\middle|\widehat{\text{CP}}(k)\right\rangle =\alpha_{j}^{\ast}\alpha
_{k}\left\langle -j\middle|\mathbb{H}\middle|-k\right\rangle \label{CP}\\
\text{CPT}  &  \text{:\ \ \ }\left\langle j\middle|\mathbb{H}%
\middle|k\right\rangle =\left\langle \widehat{\text{CP}}(k)\middle|\mathbb{H}%
\middle|\widehat{\text{CP}}(j)\right\rangle =\alpha_{k}^{\ast}\alpha
_{j}\left\langle -k\middle|\mathbb{H}\middle|-j\right\rangle \label{CPT}%
\end{align}
These symmetric properties of quantum systems are discussed in standard
monographs \cite{Symmetry1972,Symmetry1976}.

It must be noted that not all matter and antimatter states can evolve into
each other (or form a quantum superposition), since evolution of quantum
systems must preserve a number of conservative properties (such
as the electric charge). Hence only mutually convertible states, which
preserve the conservative properties, are considered here. Mutual conversions
of matter and antimatter inevitably alter the baryon numbers (for example,
conversion of a neutron into an antineutron changes this number from +1 to
-1). The possibility of the baryon number violations is known as the first
Sakharov\cite{Sakharov1967} condition of baryogenesis and is conventionally
presumed in theories dealing with the balance of matter and antimatter.

\subsection*{Asymptotic expressions for unitary evolution}

The evolution of the quantum system is governed by the Schrodinger equation%
\begin{equation}
i\frac{\partial\psi}{\partial t}=\mathbb{H}\psi,\ \ \ \ \ \ \psi
(t)=\mathbb{U}(t,t_{0})\psi(t_{0}),\ \ \ \ \ \ i\frac{\partial\mathbb{U}%
}{\partial t}=\mathbb{HU}\label{1Shr}%
\end{equation}
where $\psi(t)$ is the wave function, the Hamiltonian $\mathbb{H}(t)$ is
Hermitian, the evolution operator $\mathbb{U}(t,t_{0})$ is unitary so that its
time inverse corresponds to its conjugate transpose $\mathbb{U}(t,t_{0}%
)\mathbb{U}^{\dagger}(t,t_{0})=\mathbb{U}(t,t_{0})\mathbb{U}(t_{0},t)=I$,
where $I$ is the unit matrix.

In quantum perturbation theory\cite{Pauli1928,Dyson1949,vanHove1954,Vliet1978}%
, the time evolution operator is conventionally represented by the series
$\mathbb{U=U}_{0}+\lambda\mathbb{U}_{1}+\lambda^{2}\mathbb{U}_{2}+...$ \ that
after substitution into equation (\ref{1Shr}) yields
\begin{equation}
\mathbb{U}_{0}(t,t_{0})=\exp\left(  -i(t-t_{0})\mathbb{H}_{0}\right)
,\ \ \ \ \ \ \ \ \mathbb{U}_{l}(t,t_{0})=-i%
{\displaystyle\int\limits_{t_{0}}^{t}}
dt^{\prime}\exp\left(  -i(t-t^{\prime})\mathbb{H}_{0}\right)  \mathbb{V}%
(t^{\prime})\mathbb{U}_{l-1}(t^{\prime},t_{0})
\end{equation}
While the interaction term $\mathbb{V=V}(t)$ may depend on time, it is assumed
that the characteristic time $\tau_{v}$ associated with this change is
$\tau_{v}\sim\tau_{1}\gg\tau_{0}.$ The terms in the expansion can be easily
evaluated when $\Delta t=t-t_{0}$ is sufficiently small; specifically,
assuming that $\tau_{0}\ll\Delta t\ll\tau_{1}$ is sufficient for the
derivation of PME. In this case we write $\mathbb{U}(t,t_{0})=\mathbb{U}%
(\Delta t)$ implying that there also exists a weak dependence of
$\mathbb{U}(\Delta t)$ on $t$. Evaluation of the non-diagonal elements results
in
\begin{gather}
\left\langle j\middle|\mathbb{U}(\Delta t)\middle|k\right\rangle
\underset{j\neq k}{=}-i\lambda\left\langle j\middle|\mathbb{V}%
\middle|k\right\rangle Q_{jk}^{(1)}(\Delta t)\exp\left(  -i\frac
{(\varepsilon_{j}+\varepsilon_{k})}{2}\Delta t\right)  +...\\
\text{where\ \ \ \ \ \ }Q_{jk}^{(1)}(\Delta t)\equiv\exp\left(  -i\frac
{(\varepsilon_{j}-\varepsilon_{k})\Delta t}{2}\right)
{\displaystyle\int\limits_{0}^{\Delta t}}
dt^{\prime}\exp\left(  it^{\prime}(\varepsilon_{j}-\varepsilon_{k})\right)
=\frac{\sin\left(  \frac{(\varepsilon_{j}-\varepsilon_{k})\Delta t}{2}\right)
}{(\varepsilon_{j}-\varepsilon_{k})/2}%
\end{gather}
The diagonal elements take the form
\begin{gather}
\left\langle j\middle|\mathbb{U}(\Delta t)\middle|j\right\rangle =\exp\left(
-i\varepsilon_{j}\Delta t\right)  \left(  1-\lambda^{2}\sum_{k\neq
j}\left\langle j\middle|\mathbb{V}\middle|k\right\rangle \left\langle
k\middle|\mathbb{V}\middle|j\right\rangle Q_{jk}^{(2)}(\Delta t)+...\right) \\
\text{where\ \ \ \ \ }Q_{jk}^{(2)}(\Delta t)\equiv\frac{1+i(\varepsilon
_{j}-\varepsilon_{k})\Delta t-\exp\left(  i(\varepsilon_{j}-\varepsilon
_{k})\Delta t\right)  }{(\varepsilon_{j}-\varepsilon_{k})^{2}}%
\end{gather}
Since $Q_{jk}^{(1)}$ and $Q_{jk}^{(2)}$ are linked by the equation
\begin{equation}
D_{jk}=\frac{\left\vert Q_{jk}^{(1)}(\Delta t)\right\vert ^{2}}{\Delta
t}=\frac{2\operatorname{Re}\left(  Q_{jk}^{(2)}(\Delta t)\right)  }{\Delta
t}=2\frac{\sin^{2}\left(  \frac{(\varepsilon_{j}-\varepsilon_{k})\Delta t}%
{2}\right)  }{(\varepsilon_{j}-\varepsilon_{k})^{2}\Delta t/2}\underset{\Delta
t\rightarrow\infty}{\simeq}2\pi\delta\left(  \varepsilon_{j}-\varepsilon
_{k}\right)  , \label{Wdt}%
\end{equation}
where new quantities $D_{jk}=D_{kj}$ forming a symmetric matrix are introduced
for convenience. The derived approximations are consistent with the unitary
property
\begin{equation}
\sum_{j}\left\vert \left\langle j\middle|\mathbb{U}(\Delta
t)\middle|k\right\rangle \right\vert ^{2}=1-2\lambda^{2}\operatorname{Re}%
\left(  \left\langle k\middle|\mathbb{U}_{2}(\Delta t)\middle|k\right\rangle
\right)  +\lambda^{2}\sum_{j\neq k}\left\vert \left\langle j\middle|\mathbb{U}%
_{1}(\Delta t)\middle|k\right\rangle \right\vert ^{2}+...=1
\end{equation}
By default, the sums over indices $k$ and $j$ are evaluated over all $2n$
eigenstates. It is easy to see that, at the leading orders (up to
$O(\lambda^{2})$)$,$ the magnitudes $\left\vert \left\langle
j\middle|\mathbb{U}(\Delta t)\middle|k\right\rangle \right\vert ^{2}$ form a
symmetric matrix since $\mathbb{U}_{1}\sim\mathbb{V}$ is Hermitian. This
property, however, is not valid at the higher orders: generally, the matrix
$\left\vert \left\langle j\middle|\mathbb{U}(\Delta t)\middle|k\right\rangle
\right\vert ^{2}$ is not symmetric.

While the asymptotic representations of $\mathbb{U}(\Delta t)$ given above are
universal, these representations do not define uniquely the form of the master
equation, which depends on additional assumptions. Large systems are subject
to the process of decoherence, whose properties determine probabilistic
behaviour of the system.

\section*{Different forms of the Pauli master equation}

This section derives two alternative forms of Pauli master equation (PME):
conventional symmetric and non-conventional antisymmetric. In the next section
these forms will be shown to correspond to the symmetric and antisymmetric
extensions of thermodynamics. PME was originally suggested by Pauli
\cite{Pauli1928}, and has been repeatedly re-derived using varying techniques
and assumptions \cite{vanHove1954,Vliet1978,Joos2003,Zeh2007}. The original
approach developed by Pauli is most suitable for the derivations in this
section for a number of reasons. First, Pauli's approach explicitly
discriminates the directions of time by repeated application of decoherence,
which corresponds to setting the state of random phases at the beginning of
many sequential time intervals. Since human intuition is deeply linked to
inequality of directions of time it is common to introduce discrimination of
directions of time implicitly by implying "good" initial conditions. As
remarked by Price \cite{PriceBook}, this implicit treatment is not desirable
in applications, where the direction of time needs to be analysed and not
postulated a priori. Second, Pauli approach is based on wave functions, which
seem to be more convenient for the present analysis, which involves multi-time
correlations, than density matrices.

\subsection*{Symmetric PME}

According to Pauli's approach to the master equation, the decoherence events
occur at the times $t_{0},t_{1},...,t_{\beta},...,t_{e},$ which, as
illustrated in Figure 1, are spaced by the characteristic decoherence time
$\left(  t_{\beta+1}-t_{\beta}\right)  \sim\tau_{d}.$ The characteristic
decoherence time $\tau_{d}$ is presumed to satisfy
\begin{equation}
\tau_{0}\ll\tau_{d}\ll\tau_{1} \label{tau3}%
\end{equation}
for reasons discussed below. In the symmetric case decoherence occurs forward
in time for all (matter and antimatter) states. The energy eigenstates form
the preferred basis for decoherence: the phase of a decohered eigenstate
becomes independent of the rest of the distribution. The effect of the
decoherence on the density matrix $\mathbf{\rho}(t)$ is removing all
non-diagonal elements (as specified by the Zwanzig projection operator
\cite{Zwanzig1970}):
\begin{equation}%
\begin{bmatrix}
\rho_{-n,-n} & \vdots & \rho_{-n,n}\\
\cdots & \ddots & \cdots\\
\rho_{n,-n} & \vdots & \rho_{n,n}%
\end{bmatrix}
\ \ \ \ \underset{t=t_{\beta}}{\longmapsto}\ \ \ \
\begin{bmatrix}
\rho_{-n,-n} &  &
\begin{matrix}
& \\
\text{{\Large 0}} &
\end{matrix}
\\
& \ddots & \\%
\begin{matrix}
& \text{{\Large 0}}\\
&
\end{matrix}
&  & \rho_{n,n}%
\end{bmatrix}
\end{equation}
Irrespective of the previous state of the system, this corresponds to
transformation of the wave function into a mixture
\begin{equation}
\psi(t_{\beta}-0)\ \ \underset{t=t_{\beta}}{\longmapsto}\ \ \psi(t_{\beta
}+0)=\sum_{k}\Theta_{k}\psi^{(k)}(t_{\beta}+0) \label{1DecOp}%
\end{equation}
where $2n$ random phases $\Theta_{k}$ indicate\ that $\psi$ at $t=t_{\beta}+0$
is not a superposition but a mixture of $2n$ wave functions $\psi^{(k)}.$ Each
function $\psi^{(k)}$ in (\ref{1DecOp}) corresponds at $t=t_{\beta}+0$ to the
$k^{\text{th}}$ diagonal term of the density matrix $\rho_{k,k}=\left\langle
\psi^{(k)}|\psi^{(k)}\right\rangle $ and satisfies%
\begin{equation}
\left\langle j\middle|\psi^{(k)}(t_{\beta}+0)\right\rangle =\psi_{0}%
^{(k)}\delta_{kj},\ \ \ \ \ \ \ \ \psi_{0}^{(k)}=\left\vert \left\langle
k\middle|\psi(t_{\beta}-0)\right\rangle \right\vert \label{1DecIC}%
\end{equation}
Here, we use random phases $\Theta_{k}$ as notation that indicates mixed
states of quantum system \cite{KlimPhysA}. In this case, $\Theta_{k}$ can be
interpreted as special quantum states. This is not exactly the same but very
close in its measured effect to Pauli's work where phases were assumed to be
physically randomised. At the moment $t=t_{\beta}$ decoherence converts the
overall wave function $\psi$ of the system (which can be in any state, mixed
or pure, at $t<t_{\beta}$) into a mixture of $2n$ pure states corresponding to
the eigenstates of $\mathbb{H}_{0}$. The decoherence events change phases but
not the amplitudes of the wave functions. If $\psi=\Sigma_{k}\Theta_{k}%
\psi^{(k)},$ then the magnitude of $\psi$ is given by $\left\vert
\psi\right\vert =\left\langle \psi\middle|\psi\right\rangle ^{1/2}$, where
$\left\langle \psi\middle|\psi\right\rangle =%
{\textstyle\sum_{k}}
\left\langle \psi^{(k)}\middle|\psi^{(k)}\right\rangle $.

Due to linearity of the quantum evolutions governed by (\ref{1Shr}), each
function $\psi^{(k)}(t)$ evolves independently within each time interval
$t_{\beta}<t<$ $t_{\beta+1}$. Specifying $\Delta t=t_{\beta+1}-t_{\beta}$
allows us to determine
\begin{equation}
\left\langle j\middle|\psi^{(k)}(t_{\beta+1}-0)\right\rangle \underset{j\neq
k}{=}\lambda\left\langle j\middle|\mathbb{U}_{1}(\Delta
t)\middle|k\right\rangle \psi_{0}^{(k)}+...,\ \ \ \ \ \left\langle
k\middle|\psi^{(k)}(t_{\beta+1}-0)\right\rangle =\left(  1+\lambda
^{2}\left\langle k\middle|\mathbb{U}_{2}(\Delta t)\middle|k\right\rangle
\right)  \psi_{0}^{(k)}+... \label{fi_kj}%
\end{equation}
We introduce probabilities
\[
p_{j}^{(k)}(t)=\left\vert \left\langle j\middle|\psi^{(k)}(t)\right\rangle
\right\vert ^{2}%
\]
and kinetic coefficients%
\begin{equation}
w_{j}^{k}\underset{j\neq k}{=}\frac{\lambda^{2}}{\Delta t}\left\vert
\left\langle j\middle|\mathbb{U}_{1}(\Delta t)\middle|k\right\rangle
\right\vert ^{2}=\lambda^{2}\left\vert \left\langle j\middle|\mathbb{V}%
\middle|k\right\rangle \right\vert ^{2}D_{jk}\underset{j\neq k}{=}w_{k}%
^{j},\ \ \ \ \ \ \ \ \ w_{k}^{k}=2\lambda^{2}\operatorname{Re}\left(
\left\langle k\middle|\mathbb{U}_{2}(\Delta t)\middle|k\right\rangle \right)
=-\sum_{j\neq k}w_{j}^{k} \label{wkj}%
\end{equation}
where $D_{jk}$ is evaluated in (\ref{Wdt}) and Hermitian properties of
$\mathbb{U}_{1}$ and $\mathbb{V}$ are taken into account to establish that
$w_{j}^{k}=w_{k}^{j}$ in (\ref{wkj}). Here, we assume that $\tau_{0}\ll\Delta
t\ll\tau_{\Delta},$ where $\tau_{\Delta}\sim1/\Delta\varepsilon$ is
proportional to the characteristic density of quantum levels in the energy
space (the characteristic energy distance $\Delta\varepsilon$ between quantum
levels is very small in large systems). The probability change for every
$p^{(k)}$ over the interval $\Delta t=t_{\beta+1}-t_{\beta}$ is determined by
equations (\ref{1DecIC})-(\ref{wkj}) and takes the form
\begin{equation}
\frac{p_{j}^{(k)}(t_{\beta+1}-0)-p_{j}^{(k)}(t_{\beta}+0)}{\Delta t}=w_{j}%
^{k}p_{k}^{(k)}(t_{\beta}+0) \label{pk_dt}%
\end{equation}
With introduction of the overall probability,
\begin{equation}
p_{j}(t)=\left\vert \left\langle j\middle|\psi(t)\right\rangle \right\vert
^{2}=\sum_{k}p_{j}^{(k)}(t),\ \ \ \ \ \ \ \ \ \ \ \ \sum_{j}p_{j}(t)=1
\label{pj1}%
\end{equation}
and taking into account that $\Delta t$ is small, equation (\ref{pk_dt}) is
summed over all $k$ to give the Pauli master equation\cite{Pauli1928}%
\begin{equation}
\frac{dp_{j}}{dt}=\sum_{k}w_{j}^{k}p_{k}=\sum_{k}w_{j}^{k}p_{k}-\sum_{k}%
w_{k}^{j}p_{j} \label{PME}%
\end{equation}
The the right-hand side form of the equation explicitly involves formula for
$w_{k}^{k}$ in (\ref{wkj}).

\subsection*{Antisymmetric PME}

In the case of antisymmertic decoherence, the matter states decohere at the
moments $t_{0}<t_{1}<...<t_{\beta}<...<t_{e}$ but the antimatter states
recohere at the same moments (recoherence is decoherence backwards in time $t$
--- this is illustrated in Figure 2. Within every unitary evolution interval
$t_{\beta}\leq t\leq t_{\beta+1}$, the joint effect of matter decoherence at
$t=t_{\beta}$ and antimatter recoherence at $t=t_{\beta+1}$ is representation
of the wave function in the following form
\begin{equation}
\psi(t)=\sum_{k}\Theta_{k}\psi^{(k)}(t) \label{2Dec}%
\end{equation}
where $2n$ random phases $\Theta_{k}$ of the decohered values are
statistically independent from each other, while the corresponding $2n$ wave
functions $\psi^{(k)}(t)$ are subject to unitary evolution within the interval
interval $t_{\beta}\leq t\leq t_{\beta+1}$ and satisfy the boundary
conditions
\begin{align}
\left\langle \tilde{j}\middle|\ \psi^{(\tilde{k})}(t_{\beta}%
\ +0)\ \right\rangle  &  =\psi_{0}^{(\tilde{k})}\delta_{\tilde{k}\tilde{j}%
},\ \ \ \ \ \text{where\ \ }\ \ \ \psi_{0}^{(\tilde{k})}=\left\vert
\left\langle \tilde{k}\middle|\psi(t_{\beta}-0)\right\rangle \right\vert
,\label{bc1}\\
\left\langle \bar{j}\middle|\psi^{(\bar{k})}(t_{\beta+1}-0)\right\rangle  &
=\psi_{0}^{(\bar{k})}\delta_{\bar{k}\bar{j}},\ \ \ \ \ \text{where}%
\ \ \ \ \ \psi_{0}^{(\bar{k})}=\left\vert \left\langle \bar{k}\middle|\psi
(t_{\beta+1}+0)\right\rangle \right\vert ,\label{bc2}\\
\left\langle \bar{j}\middle|\psi^{(\tilde{k})}(t_{\beta+1}-0)\right\rangle  &
=0\ \ \ \ \text{and\ }\ \ \ \left\langle \tilde{j}\middle|\psi^{(\bar{k}%
)}(t_{\beta}+0)\right\rangle =0 \label{bc3}%
\end{align}
Note that the property specified by these equations cannot be expressed in
terms of the conventional single-time density matrix $\mathbf{\rho}(t),$ since
correlations at different time moments are needed in (\ref{2Dec})-(\ref{bc3}).
In general, this problem requires consideration of a two-time density matrix
(e.g. $\mathbf{\rho}(t_{1},t_{2})=\left\vert \psi(t_{1})\right\rangle
\left\langle \psi(t_{2})\right\vert $) but using wave functions seems to be
more convenient and is perfectly sufficient for our goals. Note that, unlike
in the case of symmetric decoherence, the antimatter states have coherent
components at $t=t_{\beta}+0$ just as the matter states have coherent
components at $t=t_{\beta+1}-0$.

While the formulae (\ref{1Shr})-(\ref{Wdt}) for the unitary evolution operator
$\mathbb{U}$ are the same as in the symmetric case, the wave function
$\psi(t)$, which is evaluated below, is different from (\ref{fi_kj}) due to
differences in the boundary conditions. While some of the terms remain similar
to (\ref{fi_kj}) at the leading order
\begin{equation}
\left\langle \tilde{j}\middle|\psi^{(\tilde{k})}(t_{\beta+1}-0)\right\rangle
\underset{\tilde{j}\neq\tilde{k}}{=}\lambda\left\langle \tilde{j}%
\middle|\mathbb{U}_{1}(\Delta t)\middle|\tilde{k}\right\rangle \psi
_{0}^{(\tilde{k})}+...,\ \ \ \ \left\langle \bar{j}\middle|\psi^{(\bar{k}%
)}(t_{\beta}+0)\right\rangle \underset{\bar{j}\neq\bar{k}}{=}\lambda
\left\langle \bar{j}\middle|\mathbb{U}_{1}(-\Delta t)\middle|\bar
{k}\right\rangle \psi_{0}^{(\bar{k})}+...
\end{equation}
the other terms change
\begin{align}
\left\langle \bar{j}\middle|\psi^{(\tilde{k})}(t_{\beta}+0)\right\rangle  &
=-\lambda\exp\left(  i\varepsilon_{\bar{j}}\Delta t\right)  \left\langle
\bar{j}\middle|\mathbb{U}_{1}(\Delta t)\middle|\tilde{k}\right\rangle \psi
_{0}^{(\tilde{k})}+...\\
&  =+i\lambda\exp\left(  -i\frac{(\varepsilon_{\tilde{k}}-\varepsilon_{\bar
{j}})}{2}\Delta t\right)  \left\langle \bar{j}\middle|\mathbb{V}%
\middle|\tilde{k}\right\rangle Q_{\bar{j}\tilde{k}}^{(1)}(\Delta t)\psi
_{0}^{(\tilde{k})}+...\nonumber
\end{align}%
\begin{align}
\left\langle \tilde{j}\middle|\psi^{(\bar{k})}(t_{\beta+1}-0)\right\rangle  &
=-\lambda\exp\left(  -i\varepsilon_{\tilde{j}}\Delta t\right)  \left\langle
\tilde{j}\middle|\mathbb{U}_{1}(-\Delta t)\middle|\bar{k}\right\rangle
\psi_{0}^{(\bar{k})}+...\\
&  =+i\lambda\exp\left(  i\frac{(\varepsilon_{\bar{k}}-\varepsilon_{\tilde{j}%
})}{2}\Delta t\right)  \left\langle \tilde{j}\middle|\mathbb{V}\middle|\bar
{k}\right\rangle Q_{\tilde{j}\bar{k}}^{(1)}(-\Delta t)\psi_{0}^{(\bar{k}%
)}+...\nonumber
\end{align}
to ensure compliance with the respective boundary conditions in (\ref{bc3}).
The additional exponential multipliers $\exp\left(  \pm i\varepsilon_{j}\Delta
t\right)  $ appear due to the phase change between the states $\left\vert
j\right\rangle $ taken at $t=t_{\beta}+0$ and $t=t_{\beta+1}-0$. As
previously, the real parts of the diagonal terms must be evaluated up to the
second order
\begin{align}
\left\langle \tilde{k}\middle|\psi^{(\tilde{k})}(t_{\beta+1}-0)\right\rangle
&  =\left(  1+\lambda^{2}\left\langle \tilde{k}\middle|\mathbb{U}_{2}(\Delta
t)\middle|\tilde{k}\right\rangle \right)  \psi_{0}^{(\tilde{k})}+\lambda
\sum_{\bar{j}}\left\langle \tilde{k}\middle|\mathbb{U}_{1}(\Delta
t)\middle|\bar{j}\right\rangle \left\langle \bar{j}\middle|\psi^{(\tilde{k}%
)}(t_{\beta}+0)\right\rangle +...\\
&  =\psi_{0}^{(\tilde{k})}\exp\left(  -i\varepsilon_{\tilde{k}}\Delta
t\right)  \left(  1-\Delta t\frac{\lambda^{2}}{2}\left(  \sum_{\tilde{j}%
\neq\tilde{k}}D_{\tilde{j}\tilde{k}}\left\vert \left\langle \tilde
{k}\middle|\mathbb{V}\middle|\tilde{j}\right\rangle \right\vert ^{2}%
-\sum_{\bar{j}}D_{\bar{j}\tilde{k}}\left\vert \left\langle \tilde
{k}\middle|\mathbb{V}\middle|\bar{j}\right\rangle \right\vert ^{2}\right)
\right)  +...\nonumber
\end{align}
The diagonal contributions from the antimatter states are evaluated in a
similar manner%
\begin{align}
\left\langle \bar{k}\middle|\psi^{(\bar{k})}(t_{\beta}+0)\right\rangle  &
=\left(  1+\lambda^{2}\left\langle \bar{k}\middle|\mathbb{U}_{2}(-\Delta
t)\middle|\bar{k}\right\rangle \right)  \psi_{0}^{(\bar{k})}+\lambda
\sum_{\tilde{j}}\left\langle \bar{k}\middle|\mathbb{U}_{1}(-\Delta
t)\middle|\tilde{j}\right\rangle \left\langle \tilde{j}\middle|\psi^{(\bar
{k})}(t_{\beta+1}-0)\right\rangle +...\\
&  =\psi_{0}^{(\bar{k})}\exp\left(  i\varepsilon_{\bar{k}}\Delta t\right)
\left(  1+\Delta t\frac{\lambda^{2}}{2}\left(  \sum_{\bar{j}\neq\bar{k}%
}D_{\bar{j}\bar{k}}\left\vert \left\langle \bar{k}\middle|\mathbb{V}%
\middle|\bar{j}\right\rangle \right\vert ^{2}-\sum_{\tilde{j}}D_{\tilde{j}%
\bar{k}}\left\vert \left\langle \bar{k}\middle|\mathbb{V}\middle|\tilde
{j}\right\rangle \right\vert ^{2}\right)  \right)  +...\nonumber
\end{align}
Taking squares of the the wave functions results in
\begin{equation}
\frac{p_{j}^{(k)}(t_{\beta+1})-p_{j}^{(k)}(t_{\beta})}{\Delta t}%
\underset{j\neq k}{=}C_{j}w_{j}^{k}p_{0}^{(k)},\ \ \ \ \ \ \ \ \ \ \ \ \frac
{p_{k}^{(k)}(t_{\beta+1})-p_{k}^{(k)}(t_{\beta})}{\Delta t}=-\sum_{j\neq
k}C_{j}w_{j}^{k}p_{0}^{(k)}%
\end{equation}%
\[
\text{where}\ \ \ \ \ p_{0}^{(\tilde{k})}=\left\vert \psi_{0}^{(\tilde{k}%
)}\right\vert ^{2}=p_{\tilde{k}}^{(\tilde{k})}(t_{\beta}%
+0),\ \ \ \ \ \ \ \ \ \ \ \ \ \ p_{0}^{(\bar{k})}=\left\vert \psi_{0}%
^{(\bar{k})}\right\vert ^{2}=p_{\bar{k}}^{(\bar{k})}(t_{\beta+1}-0)
\]
and the coefficients $w_{j}^{k}$ are still specified by (\ref{wkj}).
Evaluation of the sum $p(t)=\Sigma_{j}p^{(j)}(t)$, while taking into account
that $p(t_{\beta}+0)=p(t_{\beta}-0)$ is continuous (for any $t_{\beta})$\ and
that $\Delta t=\tau_{d}\ll\tau_{1}$ is small, yields the master equation
\begin{equation}
\frac{dp_{j}}{dt}=\sum_{k}W_{j}^{k}p_{k}=\sum_{k}C_{j}w_{j}^{k}p_{k}-\sum
_{k}C_{k}w_{k}^{j}p_{j},\ \ \ \ \text{where\ \ }W_{j}^{k}\underset{j\neq
k}{=}C_{j}w_{j}^{k}\text{\ }\ \ \text{and \ \ }W_{k}^{k}=-\sum_{j\neq k}%
W_{j}^{k}=-\sum_{j\neq k}C_{j}w_{j}^{k} \label{MEcpt}%
\end{equation}
As previously the coefficients of this master equation can depend on time
$W_{j}^{k}=W_{j}^{k}(t).$ The off-diagonal elements differ from those in SPME
only by their signs $W_{j}^{k}=\pm w_{j}^{k}$ for $j\neq k$ but the diagonal
elements are generally different $W_{k}^{k}\neq\pm w_{k}^{k}$. In absence of
matter the evolution of the antimatter states according to APME represents, as
expected, a time reversal of the evolution of these states according to SPME.
Note that, in addition to this expected result, the derived equation also
evaluates another, highly non-trivial statistical property --- how the matter
and antimatter states interact with each other.

\section*{Comparison of the two forms of PME}

First we note that the forms of PME, symmetric (\ref{PME}) and antisymmetric
(\ref{MEcpt}), can both be written as
\begin{equation}
\frac{dp_{j}}{dt}=\sum_{k}C_{j}^{\prime}w_{j}^{k}p_{k}-\sum_{k}C_{k}^{\prime
}w_{k}^{j}p_{j} \label{MEcpt2}%
\end{equation}
where the modified indicator-function $C_{j}^{\prime}$ is defined differently
for symmetric PME (SPME) and antisymmetric PME (APME) by%
\begin{equation}
\text{SPME: \ \ }C_{\tilde{j}}^{\prime}=+1,\ \ C_{\bar{j}}^{\prime
}=+1;\ \ \ \ \ \ \ \ \ \ \ \ \ \ \text{APME: \ \ }C_{\tilde{j}}^{\prime
}=+1,\ \ C_{\bar{j}}^{\prime}=-1
\end{equation}
This form is useful for analysis of the common features of these equations.

\subsection*{Invariant properties}

These properties can be summarised by the following proposition:

\begin{proposition}
\label{P1}If the system Hamiltonian is invariant\ (i.e. CP-invariant or
CPT-invariant or both), the symmetric Pauli master equation (SPME) is
CP-invariant and the antisymmetric Pauli master equation (APME) is CPT-invariant.
\end{proposition}

First we note that both constraints imposed on the Hamiltonian, (\ref{CP}) and
(\ref{CPT}) are the same for diagonal elements $j=k$ resulting in
$\varepsilon_{j}=\varepsilon_{-j}.$ For off-diagonal elements, the CP
invariance yields $\left\vert \left\langle j\middle|\mathbb{V}%
\middle|k\right\rangle \right\vert =\left\vert \left\langle
-j\middle|\mathbb{V}\middle|-k\right\rangle \right\vert $ so that $w_{k}%
^{j}=w_{-k}^{-j}$, while the CPT invariance yields $\left\vert \left\langle
j\middle|\mathbb{V}\middle|k\right\rangle \right\vert =\left\vert \left\langle
-k\middle|\mathbb{V}\middle|-j\right\rangle \right\vert $ so that $w_{k}%
^{j}=w_{-j}^{-k}$. Here we use the definition of $w_{k}^{j}$ by (\ref{wkj})
and take into account that $\left\vert \alpha_{j}\right\vert =1$ in (\ref{CP})
and (\ref{CPT}). Finally, the symmetry of the kinetic coefficients $w_{k}%
^{j}=w_{j}^{k}$ determines that both of the invariant properties, CP in
(\ref{CP}) and CPT in (\ref{CPT}), result in the same set of constrains
\begin{equation}
\text{CP and/or CPT: \ }\varepsilon_{j}=\varepsilon_{-j},\ \ w_{j}^{k}%
=w_{k}^{j}=w_{-k}^{-j}=w_{-j}^{-k}%
\end{equation}
While in general the constraints imposed on the interaction Hamiltonians by CP
invariance and by CPT invariance produce different evolutions of quantum
systems, this difference is not revealed in the PME when the decoherence time
is sufficiently short, as stipulated by (\ref{tau3}). The statement of the
proposition is then easily proved by substituting $-j$ for $j$ in (\ref{PME})
and $-j$ for $j$ and $-t$ for $t$ in (\ref{MEcpt}). This proposition indicates
that SPME can be referred to as the CP-invariant PME and APME can be referred
to as the CPT-invariant PME, although using these terms should not lead to
confusion of invariant properties of PME with those of the Hamiltonian (and
with symmetries of unitary evolutions corresponding to the Hamiltonian). The
same statement applies to the two possible extensions of thermodynamics.

\subsection*{Consistency with thermodynamics.}

The definition of entropy
\begin{equation}
S=-\sum_{k}C_{k}^{\prime}p_{k}\ln(p_{k}) \label{defS}%
\end{equation}
coincides with the conventional definition of entropy for SPME but changes the
signs of contributions of the antimatter states for APME. This definition of
entropy does not involve the degeneracy factors, as the summation is performed
over all quantum states (and not over different energy levels, which can be
degenerate). Note that the placement of $C_{k}^{\prime}$ in (\ref{MEcpt2}) and
(\ref{defS}) does not allow for interpretation of $C_{k}^{\prime}$ as
effective degeneracy factors. The entropy $S$ defined by (\ref{defS})
satisfies the following H-theorem:

\begin{proposition}
The entropy $S$ monotonically increases in evolution of probabilities
predicted by the Pauli master equation (both symmetric and antisymmetric),
unless the system is in equilibrium where the entropy remains constant.
\end{proposition}

The proof of the proposition is achieved by evaluating $dS/dt$ using
(\ref{MEcpt2})
\begin{align}
\frac{dS}{dt}  &  =-\sum_{k}C_{k}^{\prime}\left(  \ln(p_{k})+1\right)
\frac{\partial p_{k}}{\partial t}=\sum_{k,j}w_{j}^{k}\ln(p_{k})(C_{k}^{\prime
}C_{j}^{\prime}p_{k}-p_{j})=\sum_{k,j}C_{j}^{\prime}w_{j}^{k}\ln(p_{k}%
)(C_{k}^{\prime}p_{k}-C_{j}^{\prime}p_{j})\nonumber\\
&  =\frac{1}{2}\sum_{k,j}C_{j}^{\prime}w_{j}^{k}\ln(p_{k})(C_{k}^{\prime}%
p_{k}-C_{j}^{\prime}p_{j})+\frac{1}{2}\sum_{k,j}C_{k}^{\prime}w_{j}^{k}%
\ln(p_{j})(C_{j}^{\prime}p_{j}-C_{k}^{\prime}p_{k})\nonumber\\
&  =\frac{1}{2}\sum_{k,j}w_{j}^{k}\left(  C_{j}^{\prime}\ln(p_{k}%
)-C_{k}^{\prime}\ln(p_{j})\right)  (C_{k}^{\prime}p_{k}-C_{j}^{\prime}%
p_{j})\geq0 \label{H-th2}%
\end{align}
Here we use the normalisation of $p_{j}$ in (\ref{pj1}), equivalence of the
summation indices $j$ and $k$ as well as the symmetry of the coefficients
$w_{k}^{j}=w_{j}^{k}$ defined in (\ref{wkj}). We also note that $(C_{j}%
^{\prime})^{2}=1$ and $1/C_{j}^{\prime}=C_{j}^{\prime}$. Each term ($j,k$) in
the last sum is no less than zero --- this can be easily seen by considering
two alternatives $C_{j}^{\prime}=C_{k}^{\prime}$ and $C_{j}^{\prime}%
=-C_{k}^{\prime}$. The symmetric version of this H-theorem (i.e. all
$C_{j}^{\prime}=1$), which is conventional\cite{Pauli1928}, also requires that
the kinetic coefficients are symmetric $w_{j}^{k}=w_{k}^{j}$. The proof given
here is suitable for both SPME and APME.

The essential property of PME is consistency with thermodynamics as stipulated
in the following proposition.

\begin{proposition}
Pauli master equations (PME) are consistent with thermodynamics: symmetric PME
corresponds to symmetric extension of thermodynamics from matter to antimatter
and antisymmetric PME corresponds to the antisymmetric extension of
thermodynamics from matter to antimatter. Specifically, this consistency implies:

\begin{enumerate}
\item Preserving the overall probability $d\Sigma_{j}p_{j}/dt=0;$

\item Preserving the overall energy $dE/dt=0$, \ $E\equiv\Sigma_{j}%
\varepsilon_{j}p_{j};$

\item Entropy definitions that are consistent with the corresponding versions
of thermodynamics and obey $dS/dt\geq0;$

\item Symmetry of the kinetic coefficients $w_{j}^{k}=w_{k}^{j}$ and
$\left\vert W_{j}^{k}\right\vert =\left\vert W_{k}^{j}\right\vert $.
\end{enumerate}
\end{proposition}

The first property directly follows from PME (\ref{MEcpt2}). The second
property is valid since $w_{j}^{k}=w_{k}^{j}\neq0$ only if $\varepsilon
_{k}=\varepsilon_{j}$ in (\ref{wkj}), which is valid when decoherence does not
interfere with the main eigenstates $\tau_{0}\ll\tau_{d}.$ The states with the
same energy as the initial state are conventionally called "on-shelf states".
If $\tau_{d}$ is too small, equation (\ref{Wdt}) can allow for transitions
between states with different energies $\varepsilon_{k}\neq\varepsilon_{j}$
since $D_{jk}$ in (\ref{wkj}) deviates from $2\pi\delta\left(  \varepsilon
_{j}-\varepsilon_{k}\right)  $ or, alternatively, can freeze any evolution of
the system since the change in probabilities is proportional to $\Delta t^{2}$
when $\tau_{d}$\ is very small (these are quantum anti-Zeno and Zeno effects
-- see \cite{Joos2003}). The third property: the H-theorem is proven above
while the definition of entropy in (\ref{defS}) can be rewritten as
\begin{equation}
S=S_{m}\pm S_{a},\ \ \ \ \ \ \ \text{where \ \ \ }S_{m}=\sum_{\tilde{k}%
}p_{\tilde{k}}\ln(p_{\tilde{k}})\ \ \ \ \text{and}\ \ \ \ S_{a}=\sum_{\bar{k}%
}p_{\bar{k}}\ln(p_{\bar{k}}) \label{SSS}%
\end{equation}
and the plus sign in (\ref{SSS}) corresponds to SPME and is consistent with
the conventional definition of entropy $S=S_{m}+S_{a}$ in the symmetric
version of thermodynamics. The minus sign corresponds to APME and matches the
definition of apparent entropy in the antisymmetric version of thermodynamics
\cite{KM2014}, where intrinsic entropies of matter $S_{m}$ and antimatter
$S_{a}$ are summed up with the opposite signs $S=S_{m}-S_{a}.$

The symmetry of the kinetic coefficients $w_{j}^{k}=w_{k}^{j}$ reflects the
principle of detailed balance and follows from (\ref{wkj}), which is valid
when the decoherence time is sufficiently short $\tau_{d}\ll\tau_{1}$ (note
that the matrix $\left\vert \left\langle j\middle|\mathbb{U}(\Delta
t)\middle|k\right\rangle \right\vert $ is generally not symmetric when $\Delta
t$ is large). The condition $\left\vert W_{j}^{k}\right\vert =\left\vert
W_{k}^{j}\right\vert $\ follows from $w_{j}^{k}=w_{k}^{j}$ and (\ref{MEcpt}).
The equilibrium distribution achieved by SPME corresponds to equal
probabilities of all interacting states (i.e. to the microcanonical
distribution). As the size of a quantum system increases, the decoherence time
is expected to decrease, becoming very small for macroscopic objects --- this
makes the system behaviour consistent with thermodynamics. Note that the
kinetic coefficients $w_{j}^{k}$ do not depend on the decoherence time
$\tau_{d}$ as long as $\tau_{d}$ stays within its expected physical
range$\ \tau_{0}\ll\tau_{d}\ll\tau_{1}.$ Since not much is known about the
exact values of decoherence time, independence of $\tau_{d}$ adds robustness
to PME. However, an increase of decoherence time in (\ref{tau3}) towards
$\tau_{d}\sim\tau_{1}$ would lead to the need of evaluating higher order terms
in the expansion for $\mathbb{U}$, compromising the symmetry of the kinetic
coefficients. This would represent a thermodynamic violation, at least, due to
violating detailed balance in the microcanonical distributions.

\subsection*{The beginning of time and the end of time}

A solution of SPME can be extended forward in time and backward in time. While
this can be done forward in time without encountering any problems, the
backward extension has to be terminated as soon as the probability of one of
the interacting states becomes zero, otherwise the probabilities predicted by
SPME become negative, which is unphysical. This event, when the solution
cannot be extended further into the past, is called the beginning of time.
Physically, the beginning of time means that either the system is subject to
external influence (such as setting the initial conditions) that makes the
governing equations invalid or interactions of the states are terminated
$\mathbb{H}_{1}=0$ and probabilities become frozen in time. If the direction
of thermodynamic time is reversed, the system should experience the end of
time instead of the beginning of time events. Since APME runs thermodynamic
time in opposite directions for matter and antimatter, it is clear that APME
can experience both types of events, the beginning of time and the end of
time. Physically these events mean terminations of the interactions between
states and/or external interferences.

\subsection*{Interactions between the matter states and antimatter states}

While the interactions within antimatter states can be easily determined due
to the reversed-time similarity with interactions of the matter states, it is
the interaction between the matter and antimatter states that is the most
interesting and non-trivial question to be answered by PME. These interactions
can be illustrated by a simple system that has only two quantum energy
eigenstates: the matter state $j=+1$ and the antimatter state $j=-1$. The PME
for this system take the forms
\begin{equation}
\text{SPME:\ \ }\frac{d}{dt}%
\begin{bmatrix}
p_{+1}\\
p_{-1}%
\end{bmatrix}
=%
\begin{bmatrix}
-w & +w\\
+w & -w
\end{bmatrix}%
\begin{bmatrix}
p_{+1}\\
p_{-1}%
\end{bmatrix}
,\ \ \ \ \ \ \ \ \ \ \ \ \ \text{APME:\ \ }\frac{d}{dt}%
\begin{bmatrix}
p_{+1}\\
p_{-1}%
\end{bmatrix}
=%
\begin{bmatrix}
+w & +w\\
-w & -w
\end{bmatrix}%
\begin{bmatrix}
p_{+1}\\
p_{-1}%
\end{bmatrix}
\label{2ME-ma}%
\end{equation}
where $p_{+1}+p_{-1}=1$ and evolution is determined by a single kinetic
coefficient $w=w(t)$. We assume that $w>0,$ i.e. the superselection rules
allow for conversion of matter into antimatter and back. If $p_{+1}=p_{+1}%
^{0}$ and $p_{-1}=p_{-1}^{0}=1-p_{+1}^{0}$ at the initial moment $t=t_{0},$
the solutions of equations (\ref{2ME-ma}) are given by
\begin{equation}
\text{SPME:\ \ }%
\begin{bmatrix}
p_{+1}\\
p_{-1}%
\end{bmatrix}
=\frac{1}{2}%
\begin{bmatrix}
1+\left(  p_{+1}^{0}-p_{-1}^{0}\right)  \exp\left(  -2\Omega(t,t_{0})\right)
\\
1+\left(  p_{-1}^{0}-p_{+1}^{0}\right)  \exp\left(  -2\Omega(t,t_{0})\right)
\end{bmatrix}
,\ \ \ \ \ \ \ \ \text{APME:\ \ }%
\begin{bmatrix}
p_{+1}\\
p_{-1}%
\end{bmatrix}
=%
\begin{bmatrix}
p_{+1}^{0}+\Omega(t,t_{0})\\
p_{-1}^{0}-\Omega(t,t_{0})
\end{bmatrix}
\label{2ME-mas}%
\end{equation}%
\begin{equation}
\text{where\ \ \ \ \ \ }\Omega(t,t_{0})=%
{\displaystyle\int\limits_{t_{0}}^{t}}
w(t^{\prime})dt^{\prime}%
\end{equation}
In many cases, $\Omega(\infty,-\infty)$ is quite small for a single
interaction event. As expected, when $\Omega$ becomes sufficiently large, the
SPME solution converges to the state with maximal entropy $p_{+1}=p_{-1}=1/2.$
That is, if applied on a large scale, this model predicts equilibration of
matter and antimatter that, at equilibrium, should be present in equal
proportions. The APME prediction is radically different --- this equation
transfers probability from the antimatter states to the matter states until
the antimatter states can no longer be present (have zero probability). On a
large scale, this model predicts that there could not be equilibrium balance
between matter and antimatter unless antimatter is fully converted into
matter. This can be summarised in form of the proposition:

\begin{proposition}
Assuming that transitions between matter and antimatter states are possible,
the symmetric form of the Pauli master equation (SPME) predicts evolution
towards equilibrium with the same probability of matter and antimatter, while
the antisymmetric form of the Pauli master equation (APME) predicts evolution 
towards complete conversion of antimatter into matter.
\end{proposition}

The evolution specified in the proposition follows from the H-theorem and the
definition of entropy by equation (\ref{defS}). The maximal value of entropy
(subject to physical constraints) is achieved 1) for SPME when $p_{j}>0$\ are
the same for all interacting on-shelf states and 2) for APME when $p_{j}>0$
are the same for all interacting on-shelf matter states and $p_{j}=0$\ for all
antimatter and remaining matter states. The statement of the proposition does
not contradict the declared similarity of matter and antimatter. In fact, both
SPME and APME\ are based on similarity of matter and antimatter but interpret
this similarity differently. In case of APME (but not SPME) this
interpretation involves a time reversal: antimatter is converted into matter
forward in time but, in reversed time, the same process converts matter into antimatter.

\section*{Discussion}

While conventional flow of time is deeply imbedded into our intuition, it was
Boltzmann\cite{Boltzmann-book} who connected the perceived direction of the
"flow of time" with the second law of thermodynamics. He suggested that if
there was a section of Universe where entropy decreases in time, the local
population would perceive the past as the future and the future as the past in
that part of the Universe. Even now this statement would seem very strange to
many people. APME appeals to similar ideas and, for many people, may
contradict the intuitive perception of time, which makes the use of this model
more difficult. This is a disadvantage, but there are advantages associated
with APME and the antisymmetric (or CPT-invariant) approach to the
thermodynamics of antimatter. First, it connects two fundamental asymmetries
of our world---the absence of antimatter and the preferred direction of
thermodynamic time---by predicting that conversion of antimatter into matter
forward in our time is strongly favoured by thermodynamics. Demonstrating that
APME tends to convert antimatter into matter is one of the major results of
the present work (assuming that matter/antimatter conversions are allowed by
the superselection rules --- see the previous section).

The microscopic world is mostly CP-invariant. This world inevitably interacts
with thermodynamic surroundings and these interactions should also be
CP-symmetric. We see macroscopic effects of these interactions in form of
thermodynamic irreversibility but do not detect microscopic effects of these
interactions, since they do not conflict with the quantum CP invariance.
However environmental interactions that generate thermodynamic time can become
visible at a microscopic level in CP-violating systems. These interactions are
seen as apparent CPT violations even if the quantum system is strictly
CPT-preserving \cite{KlimPhysA}. Under these conditions, the ubiquitous nature
of thermodynamic interactions may lead to questioning CPT invariance. The
second advantage is that the antisymmetric approach offers a very convenient
interpretation that strictly upholds the CPT invariance: the apparent CPT
violation appears only because exact application of the CPT transformation
requires to change environmental matter into antimatter, which is practically impossible.

The system considered in the present work is subject to much stronger
thermodynamic interference than that considered in the CP-violating Kaon
decays\cite{KlimPhysA}. This strong and persistent influence of decoherence in
the derivation of PME ensures thermodynamic compliance for the evolution of
the system but suppresses the difference between CP- and CPT-invariant
Hamiltonians. It seems, however, that invariant properties of decoherence,
which remain largely unknown, should have some physical links with invariant
properties of the microscopic world. The effect of decoherence on
simulations of realistic interactions of particles and antiparticles should
involve radiation and may need to incorporate relativistic quantum mechanics,
where differences between microscopic symmetries can be more persistent.

While there are some significant advantages in considering APME and
CPT-invariant thermodynamics, these advantages do not prove that it is the
antisymmetric (and not symmetric) approach that corresponds to the real world.
Absence of antimatter in the universe may have different explanations. Would
it be possible to establish, at least in principle, which version of
thermodynamics corresponds to the real world? It seems that the answer is
generally positive but the following important points need to be taken into account.

The thermodynamic properties are not revealed in simple microscopic systems;
this requires a system of sufficient size and complexity. A simple system
placed into a thermodynamic environment does not create thermodynamic
behaviour on its own, but is subject to the thermodynamic properties of the
environment. Hence having relatively few isolated antistates or placing these
antistates into a conventional thermodynamic environment does not create an
antimatter-controlled thermodynamic system and does not solve the problem. The
challenge is to create a thermodynamic object (i.e. object that is
sufficiently complex and, conventionally, cannot be in a coherent state) that
is made not from matter but from antimatter. This object should be
sufficiently insulated from the environment so that the inter-object
interactions are overwhelmingly stronger than any environmental interference.
The difficulty of this task should not be underestimated but some encouraging
news is arriving from high-energy colliders \cite{Nature2007,H2anti2010}. It
is becoming possible to create antinuclei\cite{MuB2011} and even
antiatoms\cite{H2anti2010}, but still only in very small quantities. The other
possible object of interest is quark-gluon plasma, which can be created in
high-energy collisions -- despite its small size, this object seems to have
some thermodynamic properties \cite{Nature2007,MuB2011}. In any case, the
progress in high-energy experiments moves forward quickly and a time when the
thermodynamic properties of antimatter can be assessed experimentally may not
be too far ahead.

\section*{Conclusions}

This work introduces the antisymmetric version of the Pauli master equation
(APME). The symmetric version of the equation (SPME) is conventional. The
proved H-theorem demonstrates consistency of the symmetric and antisymmetric
versions of the PME with the symmetric and antisymmetric extensions of
thermodynamics from matter into antimatter. The symmetric versions of these
approaches are CP-invariant while the antisymmetric versions are CPT-invariant
(under conditions specified in Proposition \ref{P1}). These properties do not
necessarily correspond to, and must not be confused with, microscopic
invariant properties of the quantum Hamiltonians. The analysis of the present
work demonstrates that SPME predicts evolution towards the same probabilities
of matter and antimatter states, while APME points to full conversion of
antimatter into matter. In the absence of experimental knowledge about
thermodynamic properties of antimatter, we cannot make an experimentally
justified choice in favour of the symmetric or antisymmetric versions, however
continuing progress of high-energy physics will, hopefully, be able to resolve
this dilemma in the future.

\section*{Acknowledgement}

The author thanks Sumiyoshi Abe for insightful comments and acknowledges ARC funding.

\bigskip

\bigskip

\bigskip


\section*{Additional Information}

The author declares no competing financial interests.

\pagebreak

\bigskip \pagebreak 
\begin{figure}[h]
\begin{center}
\includegraphics[width=12cm,page=1,trim=1cm 2cm 5cm 1cm, clip ]{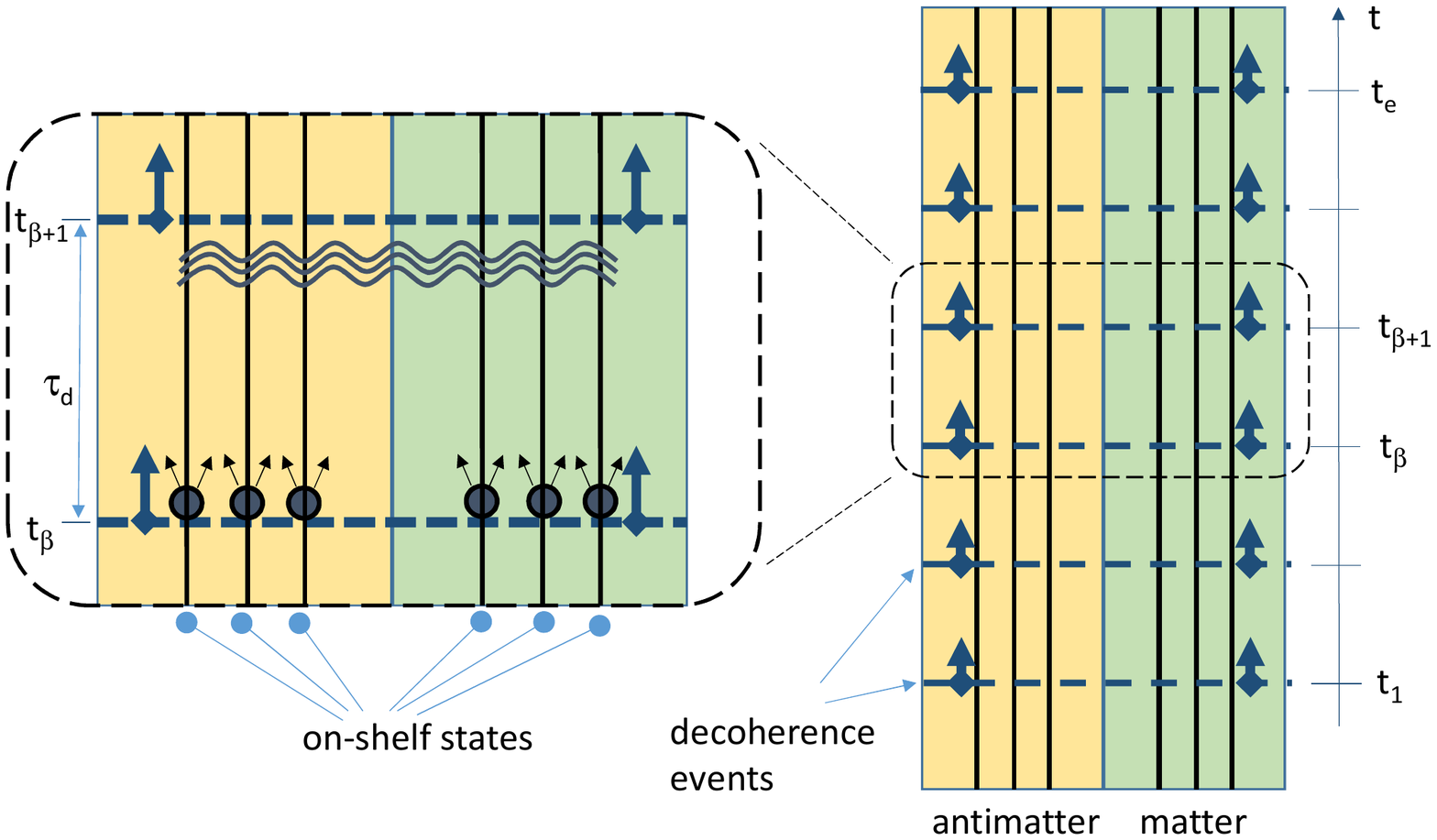}
\end{center}
\caption{Evolution of a quantum system with uni-directional (symmetric)
decoherence events separating intervals of unitary evolution. The vertical
lines show states with the same energy. The circles indicate random
phases after decoherence. Multiple waved lines indicated a mixture of wave
functions.}
\label{fig1}
\end{figure}

\begin{figure}[h!]
\begin{center}
\includegraphics[width=12cm,page=2,trim=1cm 2cm 5cm 1cm, clip ]{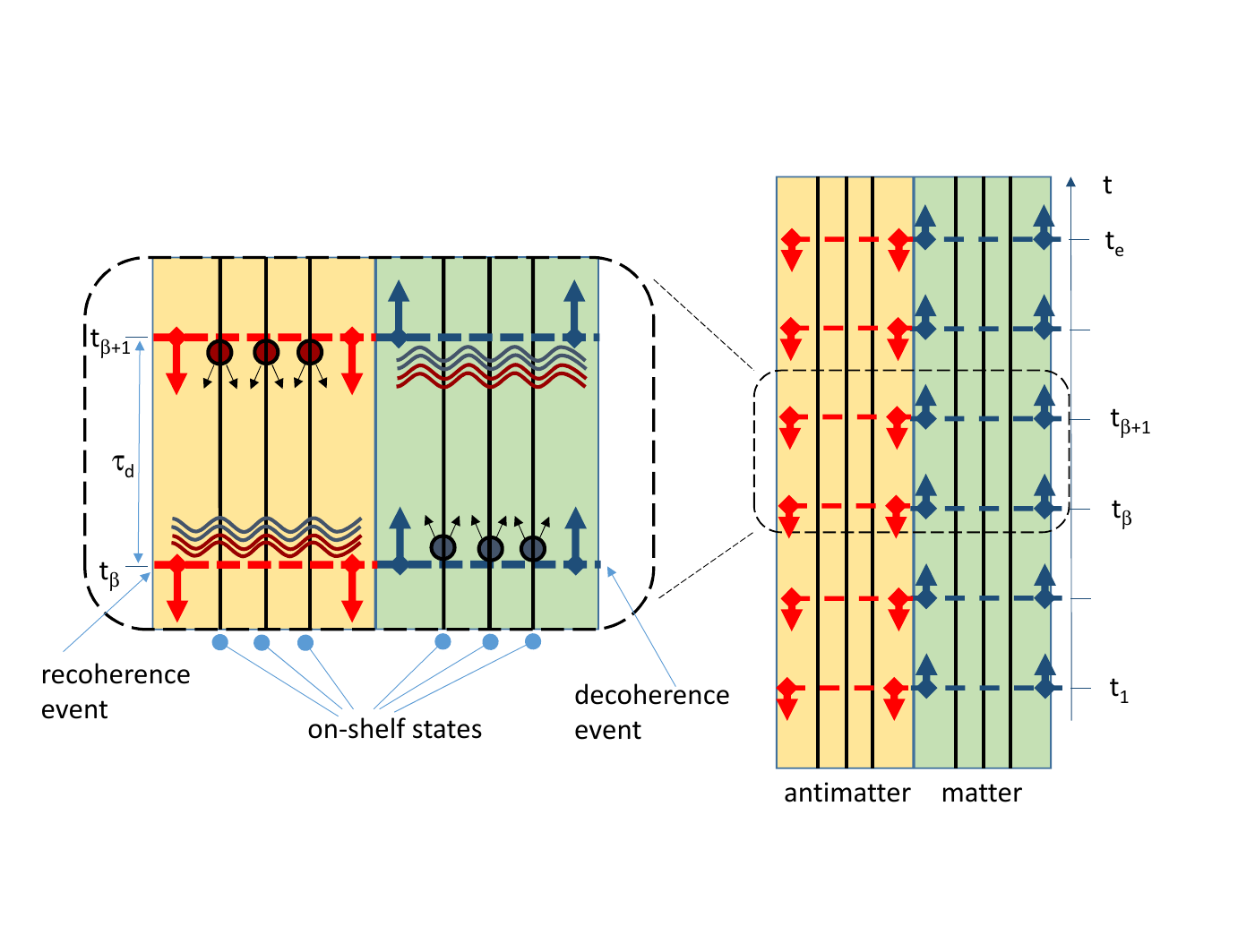}
\end{center}
\caption{Unitary evolution of a quantum system with counter-directional
(antisymmetric) decoherence events (foward in time for matter states on the
right-hand side and backward in time for antimatter states on the left-hand
side). Notations are similar to Figure 1. }
\label{fig2}
\end{figure}

\bigskip

\end{document}